\documentclass[fleqn,usenatbib]{mnras}

\usepackage[T1]{fontenc}

\DeclareRobustCommand{\VAN}[3]{#2}
\let\VANthebibliography\thebibliography
\def\thebibliography{\DeclareRobustCommand{\VAN}[3]{##3}\VANthebibliography}

\hypersetup{linkcolor=blue,citecolor=blue,filecolor=cyan,urlcolor=magenta}
\usepackage{graphicx}   % Including figure files
\usepackage{amsmath}    % Advanced maths commands
\usepackage{amssymb}    % Extra maths symbols
\usepackage{newtxtext,newtxmath}
\usepackage{color}
\usepackage{threeparttable}
\usepackage{hyperref}

\definecolor{blazeorange}{rgb}{1.0, 0.4, 0.0}
\definecolor{seagreen}{rgb}{0.18, 0.55, 0.34}
\definecolor{rufous}{rgb}{0.66, 0.11, 0.03}
\definecolor{royalfuchsia}{rgb}{0.79, 0.17, 0.57}
\definecolor{scarlet}{rgb}{1.0, 0.13, 0.0}
\definecolor{royalpurple}{rgb}{0.47, 0.32, 0.66}

\newcommand{\lta}{\lower 2pt \hbox{$\, \buildrel {\scriptstyle <}\over {\scriptstyle \sim}\,$}}
\title[Alfv\'{e}n waves]{Propagation of Alfv\'{e}n waves in the charge starvation regime}

\author[Kumar, Gill \& Lu]{
Pawan Kumar$^1$\thanks{pk@astro.as.utexas.edu}, Ramandeep Gill$^{2,3}$\thanks{ r.gill@irya.unam.mx}, Wenbin Lu$^{4,5}$\thanks{wenbinlu@berkeley.edu} \\ \\
$^1$Department of Astronomy, University of Texas at Austin, Austin, TX 78712, USA \\
  $^2$Instituto de Radioastronom\'ia y Astrof\'isica, Universidad Nacional Aut\'onoma de M\'exico, Antigua Carretera a P\'atzcuaro \# 8701, \\
~~~~~~~~Ex-Hda. San Jos\'e de la Huerta, Morelia, Michoac\'an, M\'exico C.P. 58089 \\
  $^3$Astrophysics Research Center of the Open University (ARCO), The Open University of Israel, P.O Box 808, Ra’anana 4353701, Israel\\
$^4$Departments of Astronomy and Theoretical Astrophysics Center, UC Berkeley, Berkeley, CA 94720, USA\\
$^5$Department of Astrophysical Sciences, Princeton University, Princeton, NJ 08544, USA\\
}

\begin{document}
\label{firstpage}
\pagerange{\pageref{firstpage}--\pageref{lastpage}}
\maketitle

% Abstract of the paper
\begin{abstract}

We present numerical simulation results for the propagation of Alfv\'{e}n waves in the charge starvation regime. 
This is the regime where the plasma density is below the critical value required to supply the current for the wave. 
We analyze a conservative scenario where Alfv\'{e}n waves pick up charges from the region where the charge 
density exceeds the critical value and advect them along at a high Lorentz factor. The system consisting of the Alfv\'{e}n wave and charges being carried with it, which we call charge-carrying Alfv\'{e}n wave (CC-AW), moves through a medium with small, but non-zero, plasma density. 
We find that the interaction between CC-AW and the stationary medium 
has a 2-stream like instability which leads to the emergence of a strong electric field along the direction of the unperturbed magnetic field. The growth rate of this instability is of order the plasma frequency of the medium encountered by the CC-AW. Our numerical code follows the system for hundreds of wave periods.
%when the electric field reaches an amplitude such that the stationary particles encountered by the CC-AW are accelerated to about 10\% the speed of light. 
The numerical calculations suggest that the final strength of the electric field is of order a few percent of the Alfv\'{e}n wave amplitude. Little radiation is produced by the sinusoidally oscillating currents associated with the instability during the linear growth phase. However, in the nonlinear phase, the fluctuating current density produces strong EM radiation near the plasma frequency and limits the growth of the instability.

\end{abstract}

\begin{keywords}
fast radio bursts -- stars: neutron -- radio continuum: transients
\end{keywords}

%%%%%%%%%%%%%%%%%%%%%%%%%%%%%%%%%%%%%%%%%%%%%%%%%%

%%%%%%%%%%%%%%%%% BODY OF PAPER %%%%%%%%%%%%%%%%%%

%\vspace{0.7cm}

\section{Introduction}

Finite amplitude magnetic field disturbances, or Alfv\'{e}n waves, are ubiquitous in astrophysical plasma. They are found in the interstellar medium, stellar atmospheres, accretion disks, the magnetosphere of neutron stars, etc. There is a non-zero current density along the direction of the unperturbed magnetic field ({\bf B}$_0$) when the wave-vector of an Alfv\'{e}n wave is not exactly aligned with {\bf B}$_0$. If the wave propagates in a medium of decreasing plasma density, under some generic conditions, it might face the situation where it enters a region with too little density to be able to supply the current for the Alfv\'{e}n wave even when charged particles move at the speed of light, i.e. $|\hat{\bf B}_0\cdot\nabla\times{\bf B}| > 4\pi q n$, where $\hat{\bf B}_0$ is a unit vector along ${\bf B}_0$, $n$ is the total charge density of electrons and positrons and $q$ is the elementary charge. 

The Alfv\'{e}n wave is said to have entered the charge starvation region in this case or that the wave has become charge starved. We are interested in understanding what happens to the Alfv\'{e}n wave and its interaction with particles it encounters in this condition. The application we have in mind is the magnetosphere of a neutron star where some disturbance in the interior of the star propagates to the surface and shakes up the magnetosphere. Numerous papers have suggested that some fraction of the Alfv\'{e}n wave energy when it enters the charge starvation regime is converted to coherent radio emission, e.g. \cite{Kumar+17,LuKumar2018,Yang&Zhang18,IokaZhang20,zhang20,KumarBosnjak2020,LuKumarZhang20,cooper&wijers2021,Wangetal2021,Qu&Zhang21}. The generation of coherent radiation requires that strong electric fields develop along {\bf B}$_0$ in the charge starvation regime. There are also claims to the contrary, e.g \cite{Chen&Beloborodov2020}, that nothing interesting happens to the Alfv\'{e}n wave when it becomes charge starved. These papers suggest that the wave simply advects charge particles with it when it enters the charge starvation region so that it is never truly charge starved, and that a strong electric field along {\bf B}$_0$ never develops to accelerate charge particles to highly relativistic speeds and generate coherent radiation.

%This paper investigates properties of Alfv\'{e}n waves in the charge starvation regime using appropriately designed numerical simulations, which are described in section \ref{AW-numerics}, with the goal to help settle the question whether a strong electric field along {\bf B}$_0$ develops in this interesting regime.

In Section 2, we consider propagation of charge-carrying Alfven waves (CC-AW) in vacuum and show that even in this case, an electric field parallel to the static magnetic field slowly develops over time.

In Section 3, we consider CC-AW propagating into a low density stationary plasma, and in this case we find that  the interaction leads to rapid growth of an electric field which accelerates charge particles and leads to the dissipation of the Alfven wave (AW).

\section{Advection of charge particles by Alfv\'{e}n waves at the threshold of charge starvation}

We investigate the scenario where an Alfv\'{e}n wave (AW) packet advects charge particles with it so that it is initially not charge starved. The unperturbed magnetic field is taken to be homogeneous and pointing along the z-axis. Perturbation to the magnetic field and the $\hat z$-component of the current density are described by
\begin{equation}
\begin{split}
    {\bf B_y} & = {\bf \hat y} B_y \exp(i\psi), \\
    j_0 & = {c\over 4\pi} \left( {\bf \nabla\times B_y} \right)_z = {i c k_x B_y\over 4\pi} \exp(i\psi),
     \label{AWsol}
\end{split}
\end{equation}
where $\psi \equiv k_x x + k_z z - \omega_a t$, and $\omega_a$ is the Alfv\'{e}n wave frequency.  
The particular scenario we are considering is where the AW carries charge particles with sufficient density to supply the current $j_0$ in equation (\ref{AWsol}). These particles are taken to move with Lorentz factor (LF) $\gamma_0$ that is independent of $z$ initially. In this case, the particle number density is given by
\begin{equation}
    n(t,z) = {j_0\over q c} = {i k_x B_y\over 4\pi q} \exp(i\psi),
    \label{ne}
\end{equation}
where $q$ is the charge of $e^\pm$ advected by the AW. Strictly speaking, the charge density also depends on $x$ through the dependence of $\psi$ on $x$, 
however, it does so over much larger length scales $\sim k_x^{-1}$, and so this dependence is ignored here. According to this solution, half of the wave-packet 
advects positrons with it and the adjoining half advects electrons as shown in Fig. \ref{fig-ne-init}. These solutions satisfy the particle flux continuity equation 
and Maxwell's equations.
 
The particle flux equation
 \begin{equation}
     {\partial n\over \partial t} + {\partial v_z n\over \partial z} = 0,
 \end{equation}
is 1D because particles are confined to move along the magnetic field lines; for high particle LF $v_z\approx c$. The particle density (eq. \ref{ne}) satisfies the continuity equation, as $c k_z = \omega_a$ is the dispersion relation for the Alfv\'{e}n wave. The x-component of Ampere's law (Maxwell equation for $\nabla\times{\bf B}$) gives $E_x=B_y$; plasma current along $\hat x$ is zero as charge particles can only move along the magnetic field. The y-component of the Induction equation
\begin{equation}
    {\partial E_x\over\partial z} - {\partial E_z\over\partial x} = i\omega_a B_y/c \quad {\rm or}\quad k_z E_x = \omega_a B_y/c
\end{equation}
is satisfied as can be seen from the AW dispersion relation. The other two components of the induction equations are satisfied identically. Thus, the solution given by eqs. (\ref{AWsol}) \& (\ref{ne}) satisfies particle flux and Maxwell equations to $\mathcal{O}(B_y)$. This solution has the property that charge particles of opposite signs are completely separated spatially with positrons in the region with positive $[\nabla\times{\bf B_y}]_z$ and electrons where this curl is negative. These charge particles move along the magnetic field at high Lorentz factor and provide the current-density the AW needs along $\hat z$.

The Coulomb field along the unperturbed magnetic field due to the charge separation is best calculated in the rest frame of the charges and is given by\footnote{The various factors in the expression for the Coulomb field are as follows. The charge density in the charge comoving frame is $n/\gamma_0$. The width of the causally connected slab in the comoving frame is $\sim r/\gamma_0$, and $d\ln n/dz (r/\gamma_0)$ gives the charge density difference in the two-halves of the causally connected region; the electric field vanishes when the charge is uniformly distributed.}
\begin{equation}
   E_{z,\rm coulomb} \sim  {2 \pi q n\over \gamma_0} {d\ln n\over dz} \left({r\over\gamma_0}\right)^2\sim {2\pi j_0 k_z r^2\over c\gamma_0^3},
   \label{Ez-cou}
\end{equation}
where $r$ is the distance the wave has traveled from its launching site. This is a very weak field for large $\gamma_0$ and it is superseded by the field that arises due to the current deficit that develops with time; the current deficit develops because charge particles lag the AW slightly, even when $\gamma_0$ is large, and this lag increases with time.

It should be pointed out that acceleration of particles to high LFs might pose a problem since the Coulomb field is strong for small $\gamma_0$ and the electric field direction switches as the sign of the charge density gradient changes. Thus, only half the particles of one sign are accelerated and the other half are decelerated by this field. The only way out of this problem might be that particles are accelerated to high LF while the plasma is almost neutral, which is not the condition conducive to strong electric field and particle acceleration. So, the viability of this scenario -- charges advected by Alfven wave at high LF to supply the current the wave requires -- is uncertain. Nevertheless, we assume that such a set up is physically plausible, and proceed to investigate whether this solution is stable and how it evolves with time. Our goal is to determine whether an AW packet will advect particles to avoid charge-starvation and hence energy dissipation. We show that even in the conservative scenario where particles are perfectly advected initially, the CC-AW strongly interacts with the plasma ahead of it causing the dissipation of the AW and generation of coherent radio emission. 

%The approach we have outlined is one of the most conservative way that we can use to determine the viability of the scenario where an Alfven wave packet advects charge particles with it to avoid charge-starvation and thereby prevent the development of strong electric field along $\hat z$ and thus ensures that the AW energy does not decay to power coherent radio emission.

\subsection{Development of current deficit and emergence of electric field: the vacuum case}
\label{j-deficit-sec}
\smallskip

Let us consider a fully ionized plasma where charges of opposite signs are completely separate spatially, and the particle density is $n(z)$. The current when particles are moving with Lorentz factor (LF) $\gamma_0$ or speed $v_0$ is $j$. The maximum current that this plasma can supply is
\begin{equation}
    j_{\max} = {j c\over v_0} \approx j\left( 1 + {1\over 2\gamma_0^2}\right)\,,
\end{equation}
where $c/v_0=\beta_0^{-1}=(1-\gamma_0^{-2})^{-1/2}\approx(1+1/2\gamma_0^2)$ for $\gamma_0\gg1$. 
If the current deficit, $\delta j \equiv j_0 - j$, were to exceed $j/2\gamma_0^2$ then no amount of acceleration of plasma can make up for this current deficit as long as the plasma density does not increase ($j_0$ is given by eq. \ref{AWsol}). If plasma density were to increase at one location then that can be only at the expense of lower density at another location, thereby making current deficit larger at this other location.

Let us start with zero current deficit everywhere, i.e. $\delta j(z,t=0) = 0$. The current deficit develops with time in some regions of the AW due to the fact that particle speed is always less than the speed of Alfv\'{e}n waves. AW speed is given by: $v_a/c = 1 - \omega_p^2/2\omega_B^2$ \citep{Krall-Trivelpiece-73,kulsrud2005}, and the corresponding LF is $\gamma_A \sim \omega_B/\omega_p\gg 10^{6}$ in NS magnetosphere; where $\omega_p = (4\pi n q^2/m)^{1/2}$ is the plasma frequency, and $\omega_B = qB_0/mc$ is the cyclotron frequency. The slip that develops between particles and the AW in time $T$ is $\delta z = Tc/2\gamma_0^2$, and thus $\delta j\approx \delta z (d j_0/dz)$, or
\begin{equation}
    \delta j \approx {\rm sign}\left( {d j_0\over dz} \right) {j_0\over 2\gamma_0^2}\left( {Tc\over \lambda_a}\right),
\end{equation}
where $\lambda_a$ is AW wavelength along z. This equation tells us that the current deficit increases linearly with time, due to the finite particle speed, and in one wave period the deficit becomes too large to be eliminated entirely no matter how rapidly particles are accelerated. The z-component of the electric field that develops due to this current deficit or surplus is 
\begin{equation}
    E_z(z,t) = -4\pi \int dt\, \delta j.
\end{equation}    
Or
\begin{equation}
   E_z = 2\pi Tc \int {dz\over c}\, {1\over \gamma_0^2} {d j_0\over dz} \approx {2\pi T \over \gamma_0^2} \left[ j_0(z,t) \right]
\end{equation}
where we took $\gamma_0$ to be independent of $z$, and the last equality is obtained by integrating over half a wavelength of the AW to obtain the peak amplitude of $E_z$. Comparing this $E_z$ with the Coulomb field (eq. \ref{Ez-cou}), we see that the electric field that develops due to current deficit is larger by a factor $\gamma_0/(rk_z)$. We can rewrite $E_z$ in terms of the magnetic field perturbation associated with the AW by making use of the expression for $j_0$ (eq. \ref{AWsol})
\begin{equation}
    E_z \sim {B_y (\omega_a T)\over 2 \gamma_0^2} \left({k_x\over k_z}\right)
    \label{Ez-1}
\end{equation}
The LF of e$^\pm$ changes in time T due to this electric field by an amount
\begin{equation}
    \delta\gamma_0 \sim {q E_z c T\over m c^2} \sim {q B_y\over m c} \left[ {(\omega_a T) T\over 2\gamma_0^2} \right] \left( {k_x\over k_z} \right) 
\end{equation}
This suggests that particles might attain an asymptotic LF of 
\begin{equation}
    \gamma_0 \rightarrow \left[ { (q B_y) \omega_a T^2 k_x\over 2 m c k_z} \right]^{1/3}.
\end{equation}
However, it should be pointed out that the electric field in one-quarter of the wavelength where the gradient of $j_0$ is positive will develop negative electric field, and in the adjoining quarter wavelength the field would be along positive $z$. Therefore, $e^+$s in the former case would be decelerated while in the latter case accelerated. This velocity different would grow quadratically with time, and the simple AW solution described by equation (\ref{AWsol}) will not hold for very long. 

Moreover, what we have described in this section is a highly idealized situation where the charge carrying AW (CC-AW) is propagating through a medium with zero plasma density; from here on we shall use the compact name {\bf CC-AW} for the system consisting of an Alfv\'{e}n wave plus the particles being advected with it at high speeds.  A physically more realistic situation is that the CC-AW encounters a nearly stationary plasma of low density as it travels further out into the magnetosphere. The interactions between the Alfv\'{e}n wave, the charge particles it is advecting along at high speeds, and the stationary plasma they encounter, make this system complex and full of interesting physics. The consequences of these interactions for the development of the z-component of the electric field, and the evolution and dissipation of AWs are topics that we explore in the next two sections. 

\section{Interaction between charge carrying Alfv\'{e}n wave (CC-AW) and stationary plasma}
\label{AW-numerics}

We consider the physics of an Alfv\'{e}n wave packet that is advecting charge particles with it, and they encounter stationary electron-positron plasma of low density that is charge neutral. We shall consider a set of 2D equations (one space dimension and time), which is faster to solve numerically. The CC-AW is taken to propagate along the z-axis. The x-dependence of all variables is $\exp(i k_x x)$, and they are all independent of $y$. The conservation of magnetic flux together with $k_y=0$, $k_x\not=0$, and $k_z\not=0$ ensures that only the $\hat y$ component of the magnetic field perturbation, $B_y$, is non-zero. Thus, the non-zero components of the Maxwell's equations are 
\begin{align}
  \label{EM-eqs}
    -\partial_z B_y & = \partial_t E_x/c, \quad  i k_x B_y = 4\pi j_z/c + \partial_t E_z/c  \\
     \partial_z E_x - i k_x E_z & = -\partial_t B_y/c  
 \label{EM-eqs2}
\end{align}
These combined with the following particle continuity and momentum flux conservation equations provide a complete description of the 2D problem,
\begin{align}
    \label{particle-dyna0}
      \partial_t n + \partial_z (n v_z)  = 0 \quad \& \quad \partial_t(\gamma v_z) + v_z \partial_z(\gamma v_z) & = {q E_z\over m}  \\
       {\rm or}\quad\quad \partial_t(v_z) + v_z \partial_z(v_z) & = {q E_z\over m\gamma^3}.
      \label{particle-dynamics}
\end{align}

\begin{figure}
\centering
\includegraphics[width = 0.47\textwidth]{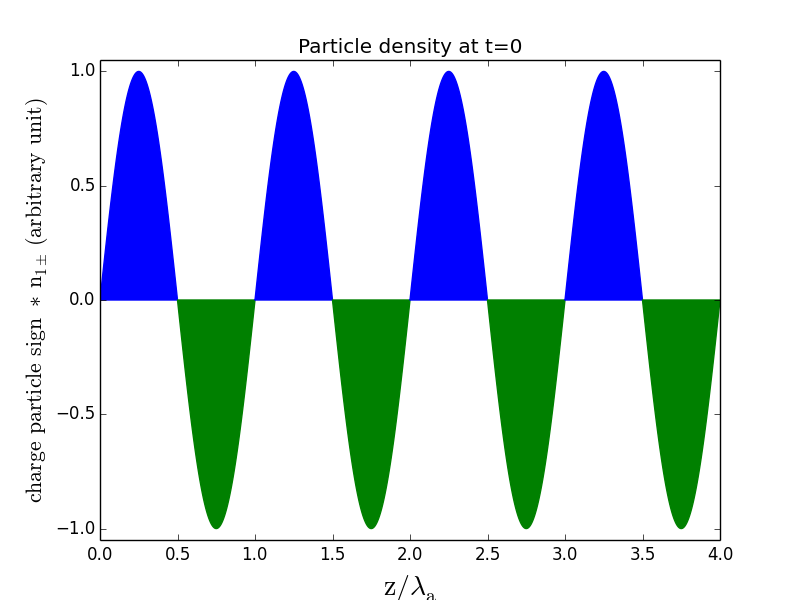}
\vskip -0.07 in

\caption{Shown here is the initial density, at $t=0$, of e$^\pm$ times the sign of the charge particle advected by the Alfv\'{e}n wave as a function of the z-coordinate which is the direction of propagation of the wave; $n_{1\pm}$ is given by eq. \ref{init-cond}, $\lambda_a \equiv 2\pi c/\omega_a$ is AW wavelength along the z-direction. Note that positron density is shown here as positive numbers and electron density as negative numbers. The charges are fully separated initially, as the different color shadings indicate, and they are all moving at the same speed v$_0$ corresponding to Lorentz factor 10$^2$ initially. The current density at $t=0$ is precisely equal to $c( \nabla\times{\bf B})_z/4\pi$ as given by equation \ref{init-cond}.}
\label{fig-ne-init}
\end{figure}

We consider only the z-component of the momentum equation as particles are locked in the lowest Landau state in the presence of the strong magnetic field being considered in this work, and hence confined to move along the magnetic field. 

The initial conditions we use are\footnote{The initial conditions for the numerical analysis presented in this work (eq. \ref{init-cond}) is motivated in part by the PIC simulations carried out by \cite{Chen&Beloborodov2020} of Alfv\'{e}n wave propagation in a stratified medium where the particle density for the first few wavelengths is sufficiently high (so that the waves are not charge starved) and then the wave enters a medium with density much smaller than the critical density for the AW. They find the development of charge separation, and acceleration of particles to Lorentz factors of a few 10s when the wave enters the low density medium. Thus, according to their simulations, the AW carries charge particles with it into vacuum thereby preventing the charge starvation. The viability and stability of this scenario when the wave travels a distance of 10$^2$s of wavelength is investigated in this work.},
\begin{equation}
  \label{init-cond}
 \begin{split}
    B_y(t=0, z) &= \epsilon B_0 \sin(\omega_a z/c), \quad {\rm for} \quad 0<z<z_0 \\
    E_x(t=0, z) &= \epsilon B_0 \sin(\omega_a z/c), \quad E_z(t=0,z) = 0, \\
    E_z(t,z=tc) &= 0, \quad j_{z_1}(t=0, z) = c\left[ \nabla\times{\bf B_y}\right]_z/4\pi \\
    v_{1z}(t=0, z) & = v_0 \equiv\left(1- \gamma_0^{-2}\right)^{1/2} \\
   \quad n_{1}(t=0, z) & = j_z/[q v_0 {\rm sign}(j_z)],
 \end{split}
\end{equation}
where $B_0$ is the unperturbed magnetic field strength, $\epsilon$ and $\omega_a$ are the AW amplitude (dimensionless) and frequency respectively, and the AW initially has only positrons in the part of the wave packet where the current density is positive, i.e. $j_z >0$, and electrons where $j_z<0$. All particles advected by the wave are taken to have speed v$_0$ initially. The initial particle density distribution along z-axis is shown in Fig. (\ref{fig-ne-init}). 

The boundary condition at the head of the AW is that stationary plasma, that is charge neutral, with prescribed constant density $n_2$ enters the wave, and $E_z$ is the Coulomb field of the CC-AW. The boundary condition at the rear end of the AW is that no particles enter the system here. 

For convenience of later use we define a frequency associated with the particle density advected with the AW
\begin{equation}
    \omega_{p_1}(z) \equiv \left[ {4\pi q^2 n_{1}(z) \over m} \right]^{1/2},
  \label{wp1}
\end{equation}
which is not the physical plasma frequency as it is missing a factor of $\gamma_0$; we will include the correct factor when using $\omega_{p_1}$ to construct the plasma frequency. 

The CC-AW -- AW along with particles with density $n_{1}(z)$ advected by the wave -- encounters stationary $e^\pm$ plasma of density $n_2$ as they travel outward to larger $z$. In all of our simulations we take $n_2$ to be independent of $t$ and $z$ before particles are accelerated by the CC-AW. And we experiment with a number of different values of $n_2/{\rm max}[n_1(z)]<1$ in our numerical simulations.

We consider in the next sub-section an approximation where the fields $B_y(t,x,z)$ and $E_x(t,x,z)$ are specified, i.e. these components of the fields are pre-set to be the time translation of the function in equation (\ref{init-cond}). This simplification makes it possible to easily understand, using analytical methods, some of the key features of our numerical simulation results. Numerical solutions of the exact problem, without these approximations, are presented in \S\ref{AW-subsec2} where we show that the main features of the solutions are essentially the same as obtained with the aforementioned approximations. 

\subsection{Charge starved Alfv\'{e}n wave propagation when the magnetic field perturbation is pre-specified}
\label{AW-subsec1}

We assume in this sub-section that the magnetic field perturbation associated with the Alfv\'{e}n wave packet shifts with time to larger $z$ at speed $c$ without any distortion to its amplitude or phase. This is not strictly correct. However, this assumption simplifies the calculation as the only EM field variable we need to evolve in this case is $E_z$; the other two field components, viz. $B_y$ and $E_x$, are known as per this assumption from the initial condition for the Alfv\'{e}n wave packet. This assumption makes it possible to obtain approximate analytic solutions for the dynamics of the CC-AW system in the charge starvation regime and obtain physical insights regarding some of its basic properties. The magnetic field perturbation, $B_y$, according to this assumption, is explicitly given by:
\begin{equation}
    B_y(t,x, z) = \epsilon B_0 \sin\left( {\omega_a z\over c} + k_x x - \omega_a t \right),
   \label{By-tz}
\end{equation}
where we have also assumed that the Lorentz factor of the Alfv\'{e}n wave is much larger than the LF of particles, and thus taking the AW dispersion relation to be $\omega_a = k_z c$ is quite accurate.

\begin{figure}
\centering
\hspace*{-0.5cm}
\includegraphics[width = 0.48\textwidth,height=7.2cm]{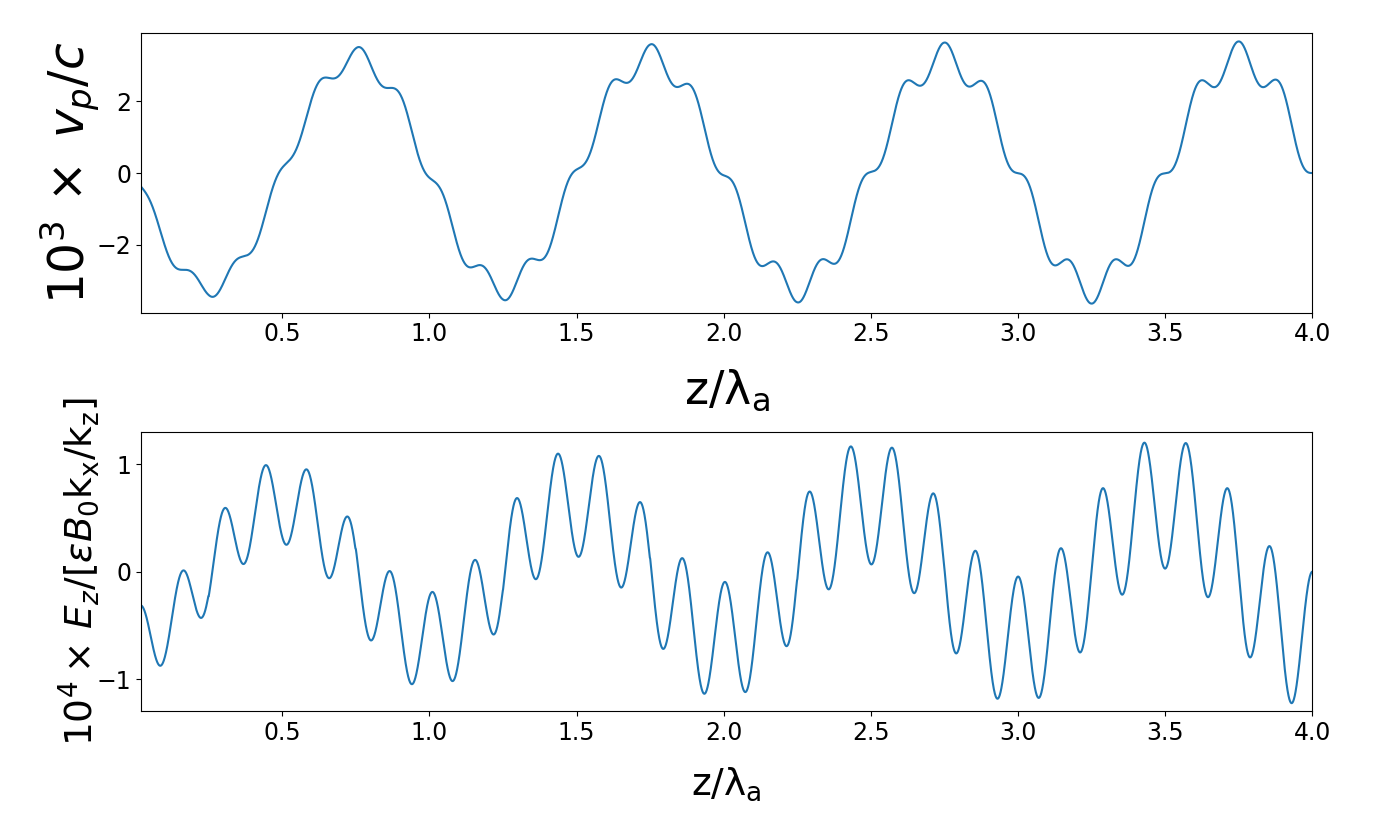}
\vskip -0.07in
\caption{Stationary $e^\pm$ encountered by charge-carrying-AW-packet (shown in Fig. \ref{fig-ne-init}) get accelerated by the z-component of the electric field ($E_z$) shown in the lower panel. The upper panel shows the speed of $e^+$s, at a fixed time, along the extent of AW packet on the z-axis ($ct$ should be added to the z-coordinate shown in the figure to obtain the lab frame location of the AW); the unperturbed magnetic field is along $z$, $\epsilon=10^{-3}$, and $\epsilon B_0$ is the amplitude of the Alfv\'{e}n wave packet. The positron speed and $E_z$ were calculated at $t=10 T_{a}$ by numerically solving eqs. \ref{EM-eqs}--\ref{particle-dynamics} with the simplifying assumption that $B_y$ is given by equation \ref{By-tz};  $T_{a}=2\pi/\omega_a$ \& $\lambda_{a} = c T_{a}$ are the Alfv\'{e}n wave period and wavelength respectively. We took $\gamma_0=10^2$, $k_x/k_z=0.5$, $\omega_B/\omega_a = qB_0(m c \omega_a) = 10^5$, the plasma frequency of the stationary $e^\pm$ plasma, $\omega_{p_2}$, is 7.1 times the Alfv\'{e}n wave frequency $\omega_a$, and the peak value of $\omega_{p_1}$ (defined by eq. \ref{wp1}) is $7.1\, \omega_a$ for these simulations. }
\label{fig-vp-Ez}
\end{figure}

\begin{figure*}
\centering
\hspace*{-0.3cm}
\includegraphics[width = 0.98\textwidth, height=13cm]{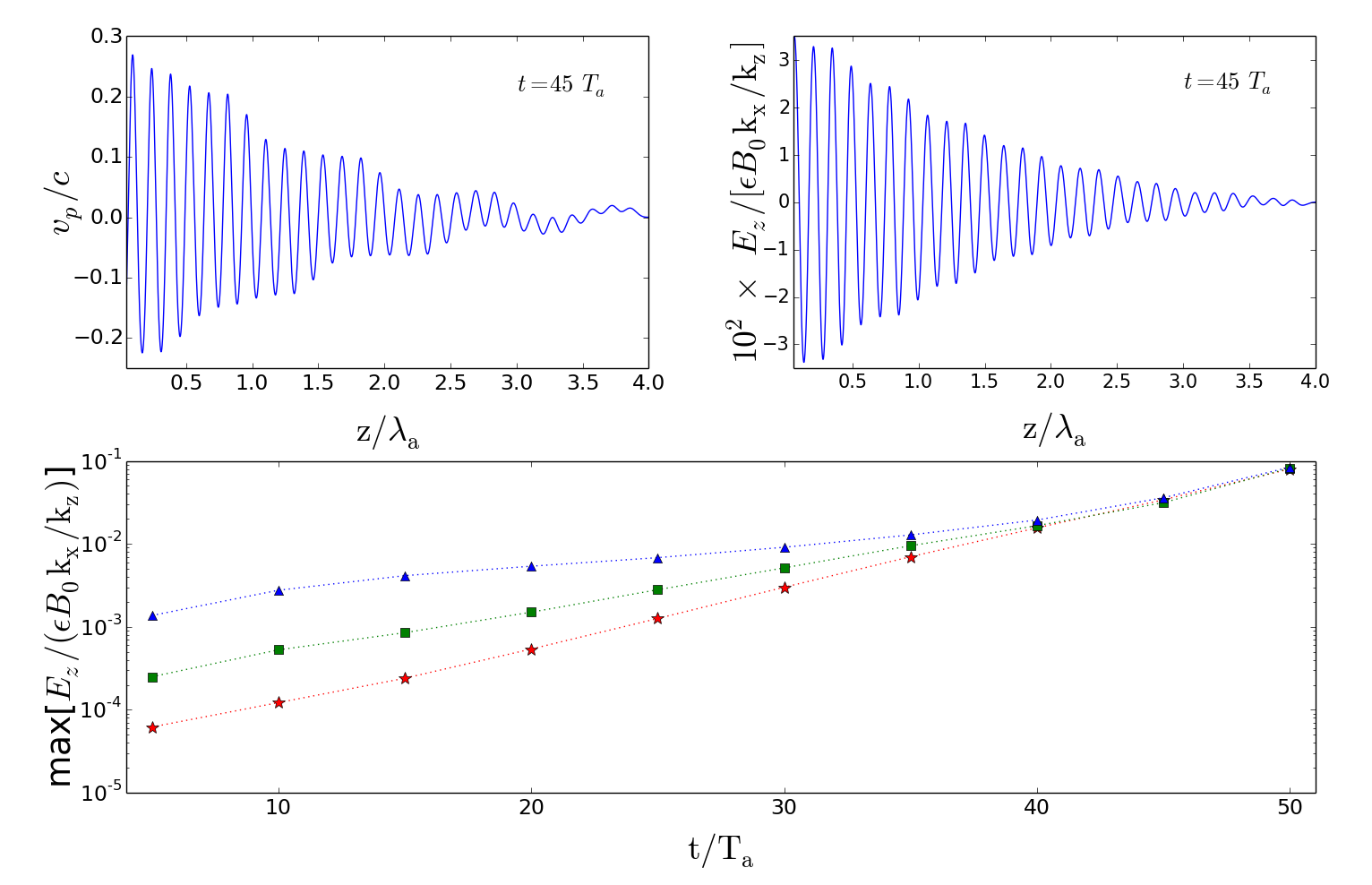}
\vskip -0.1in
\caption{The top left and right panels show $v_p$ and $E_z$ at $t=45 T_{a}$ -- Fig. \ref{fig-vp-Ez} also showed these variables but at $t=10 T_{a}$. The parameters for the simulations here are the same as in the calculations presented in Figs. \ref{fig-vp-Ez}, i.e. $\gamma_0=10^2$,   $k_x/k_z=0.5$, $\omega_B/\omega_a = qB_0/(m c \omega_a) = 10^5$, $\omega_{p_1}/\omega_a =7.1$ (at the peak of AW), and $\omega_{p_2}/\omega_a =7.1$; see the caption for Fig. \ref{fig-vp-Ez} for the explanation for what $v_p$, $E_z$, $T_a$, etc. stand for. Note that the electric field $E_z$ at $t=45 T_{a}$ (top-right) is larger by a factor $\sim 200$ compared with the field at $t=10 T_{a}$ (Fig. \ref{fig-vp-Ez}). Furthermore, the data in Fig. \ref{fig-vp-Ez} clearly shows the superposition of free oscillation at frequency $\omega_{p_2}=7.1\omega_a$ and the forced oscillation at frequency $\omega_a$. The free oscillation grows exponentially with time as described in \S\ref{AW-subsec2}, whereas the forced oscillation does not, and the result is that the oscillations of $v_p$ and $E_z$ at $t=45 T_{a}$ is dominated by frequency $\omega_{p_2}$ as seen in the upper panels of this figure. The lower panel shows the global maximum of 
$|E_z(t,z)|/(\epsilon B_0 k_x/k_z)$ -- along the finite width of the AW packet at a fixed $t$ -- as a function of time for three different values of $e^\pm$ density encountered by the AW packet -- the density $n_2/{\rm max(n_{1\pm})}$ for the lowest curve (points marked by red-stars) is 0.5, points denoted by green-squares have $n_2/{\rm max(n_{1\pm})}=0.125$, and for triangles it is 0.031; where $n_{1\pm}$ is given by eq. \ref{init-cond}, and all points in this panel are obtained from our numerical simulations of Alfv\'{e}n wave propagation with initial and boundary conditions specified by equation \ref{init-cond}. }
\label{fig-Ez2}
\end{figure*}

%The plasma density encountered by the CC-AW is taken to be constant in our numerical simulations.
We show in Figs. \ref{fig-vp-Ez} \& \ref{fig-Ez2} numerical solutions of equations \ref{EM-eqs}--\ref{particle-dynamics}, with the approximation mentioned at the beginning of this subsection, for one set of parameters. We can understand the magnitude and the general behavior of the z-component of the electric field, $E_z$, that develops in this interaction using the following approximate calculation. 

Let us write the current density as the sum of two parts. The first piece, $j_{z_1}$ is due to $e^\pm$ advected by the AW, and the second component, $j_{z_2}$, is due to the charge particles encountered by the CC-AW, which are accelerated along the $\hat z$ direction due to non-zero $E_z$. Rewriting Amp\`{e}re's law with these two components of the current density separately we find
\begin{equation}
    \left( {\bf \nabla\times B_y} \right)_z - {4\pi j_{z_1}\over c} \equiv -{4\pi j'\over c} = {4\pi j_{z_2}\over c} + {1\over c} {\partial E_z\over \partial t} 
\end{equation}
Or
\begin{equation}
    4\pi j_{z_2} + \partial_t E_z = -4\pi j'.
    \label{jz2-1}
\end{equation}
The non-zero current-density deficit, $j'$, arises from the fact that charge particles are moving at a speed $v_0$ that is smaller than the Alfv\'{e}n-wave speed, and is given by (see \S\ref{j-deficit-sec} for a detailed derivation)
\begin{equation}
  \label{j-prime}
    j'(t,z) = \delta z \partial_z j_{z_1} = -\left[{\omega_a\delta z\over c}\right] {c k_x B_y(t,z)\over 4\pi},
\end{equation}
where $B_y(t,z)$ is given by equation (\ref{By-tz}),
\begin{equation}
    \delta z \approx -{ct\over2\gamma_0^2},
\end{equation}
%\rg{Shouldn't $\delta z$ be a positive quantity? Also, Eq. 22 is already defined with a negative sign, which cancels the negative sign here.} \pk{I think the signs in eqs. 22 \& 23 are okay. The negative sign in 22 is because $\partial_z j_z\propto \partial^2_z B_y \propto -B_y$. And the negative sign in 23 reflects the fact that particles lag the Alfven wave packet.}
and $t$ is the time elapsed since the AW packet entered the charge starvation region.

The current density due to the $e^\pm$ encountered by the CC-AW, which are accelerated to speed $v_\pm$ is
\begin{equation}
    j_{z_2} = q n_{2+} v_{2+} - q n_{2-} v_{2-} \equiv j_{2+} + j_{2-} = \sum_\pm j_{2\pm},
\end{equation}
and the time derivative of this current density is
\begin{eqnarray}
    {d j_{z_2}\over dt} &=& \sum_\pm q_\pm\left[ n_{2\pm}{d v_{2\pm} \over dt}
       + v_{2\pm} {d n_{2\pm}\over dt} \right] \\
       &=& \sum_\pm \left[ {q^2 E_z n_{2\pm} \over m \gamma_{2\pm}^3} - j_{2\pm} {\partial v_{2\pm}\over \partial z}\right],
       \label{jz2-2}
\end{eqnarray}
where $q_\pm \equiv \pm q$, and $dj_\pm/dt \equiv \partial_t j_\pm + v_\pm\partial_z j_\pm$. We have made use of particle conservation equation, and the equation of motion of a charge particle under electric field, i.e. $d (\gamma_\pm v_\pm)/dt = q_\pm E_z/m$ or $dv_\pm/dt = q_\pm E_z/(m\gamma_\pm^3)$ in arriving at the final expression in equation (\ref{jz2-2}). It follows from equation (\ref{jz2-2}) that
\begin{equation}
   {\partial j_{z_2}\over \partial t} = {\omega_{p_2}^2 E_z\over 4\pi\gamma_2^3} - \sum_\pm {\partial (v_{2\pm} j_{2\pm})\over \partial z},
   \label{jz2-t-exact}
\end{equation}
where 
\begin{equation}
    \omega_{p_2}^2 \equiv {4\pi q^2 n_2\over m}, \quad {1\over \gamma_2^3} \equiv \sum_\pm {n_2^\pm\over n_2\gamma_{2\pm}^3} \quad{\rm \&} \quad n_2\equiv n_{2-} + n_{2+}.
  \label{wp2}
\end{equation}

The second term in the expression for $\partial_t j_{z_2}$ vanishes when electrons and positrons have opposite velocities, $v_{2-}(t,z) = - v_{2+}(t,z)$, as one might expect when the electric field reverses direction on a length scale much smaller than the AW wavelength, i.e. $\omega_{p_2} \gg\omega_a$; we note that this approximation breaks down when $|v_{2\pm}|/c \gtrsim 10^{-1}$. We will use this approximation in our derivation to obtain insights in the behavior of the system. The equation for $j_{z_2}$ reduces to
\begin{equation}
   {\partial j_{z_2}\over \partial t} = {\omega_{p_2}^2 E_z\over 4\pi\gamma_2^3}
   \label{jz2-t}
\end{equation}
with this approximation. Finally, we take the time derivative, at fixed $z$, of equation (\ref{jz2-1}) and make use of (\ref{jz2-t}) to arrive at
\begin{equation}
    \partial^2_t E_z + {\omega_{p_2}^2 E_z \over \gamma_2^3} \approx -4\pi \partial_t j',
    \label{Ez-eq1}
\end{equation}
where $j'$ is given by equation (\ref{j-prime}). The oscillator equation for $E_z$ can be solved exactly when $\gamma_2\approx 1$, and the solution is given by 
\begin{equation}
    E_z(t,z) \approx A \sin(\omega_{p_2} t + \phi) - {4\pi \partial_t j'\over \omega_{p_2}^2},
    \label{Ez-2}
\end{equation}
where we have taken $\omega_{p_2} \gg\omega_a$. The constant $A$ is determined by the condition that at a fixed $z$, $E_z(t_0,z)=0$ at time $t_0 = (z-z_0)/c$ that is when the AW head arrives at that point; where $z_0$ is the z coordinate at the head of the AW wave at $t=0$.  This gives
\begin{equation}
    A = {4\pi \partial_t j'(z,t)|_{t_0}\over \omega_{p_2}^2 \sin(\omega_{p_2} t_0 + \phi) },
    \label{amp-1}
\end{equation}
the phase $\phi$ can be determined by the condition that at time $t_0$, at $z$, $j_{z_2}=0$, and therefore, $\partial_t E_z = -4\pi j'(t_0,z)$ as per equation (\ref{jz2-1}). 

The solution obtained above for the electric field (eqs. \ref{Ez-2} \& \ref{amp-1}) is a superposition of free and forced oscillations with frequencies $\omega_{p_2}$ and $\omega_a$ respectively. We see this behavior very clearly in our numerical simulation results presented in Fig. \ref{fig-vp-Ez}. The analytic solution (eq. \ref{Ez-2}) breaks down when particles encountered by the CC-AW are accelerated to speed $\gtrsim c/10$. The second term in equation (\ref{jz2-t-exact}) we neglected is no longer small in that case, and has to be included in the calculation. Moreover, it is no longer a valid approximation to neglect the action of $E_z$ on the current density associated with charge particles advected by the Alfven wave packet ($j_{z_1}$), as we did in our calculations above. Our numerical simulations, of course, keep track of all these effects.

%and the result is that the free oscillation component of the electric field grows exponentially with time and dominates the forced oscillation term in equationhttps://www.overleaf.com/project/60e738a55a93b46524bad070 (\ref{https://www.overleaf.com/project/60e738a55a93b46524bad070Ez-2}) after a while. This is shown in figure (\ref{fig-Ez2}).

The electric field strength, shown at $t=10 T_a$ in Fig. \ref{fig-vp-Ez}, is weaker than given in equation (\ref{Ez-1}) by a factor $\sim \omega_{p_2}/\omega_a$; [$T_a\equiv 2\pi/\omega_a$ is the Alfv\'{e}n wave period]. This is because the fresh plasma encountered by the CC-AW shields the electric field effectively but not completely; the residual field is substantially smaller than the field strength generated by current deficit of the CC-AW system as given in eq. \ref{Ez-1}.

However, the electric field grows exponentially with time as shown in the lower panel of Fig. \ref{fig-Ez2}; $E_z$ at $45 T_a$ is larger by a factor $\sim 500$ than the field at $10 T_a$, and exceeds the field strength given by eq. \ref{Ez-1}. We have tried different sets of initial conditions for $B_y$ than the one given in equation (\ref{init-cond}), and they all have exponentially growing $E_z$; the memory of the initial condition is lost after a few or few-tens of AW periods depending on $\gamma_0$.
The exponential amplification of $E_z$ is a result of an instability that is very similar to the well known two-stream instability. The linear stability analysis of the CC-AW system encountering low density plasma is presented below in \S\ref{instability}.

\begin{figure*}
\centering
\hspace*{-0.5cm}
\includegraphics[width = 0.93\textwidth,height=12.5cm]{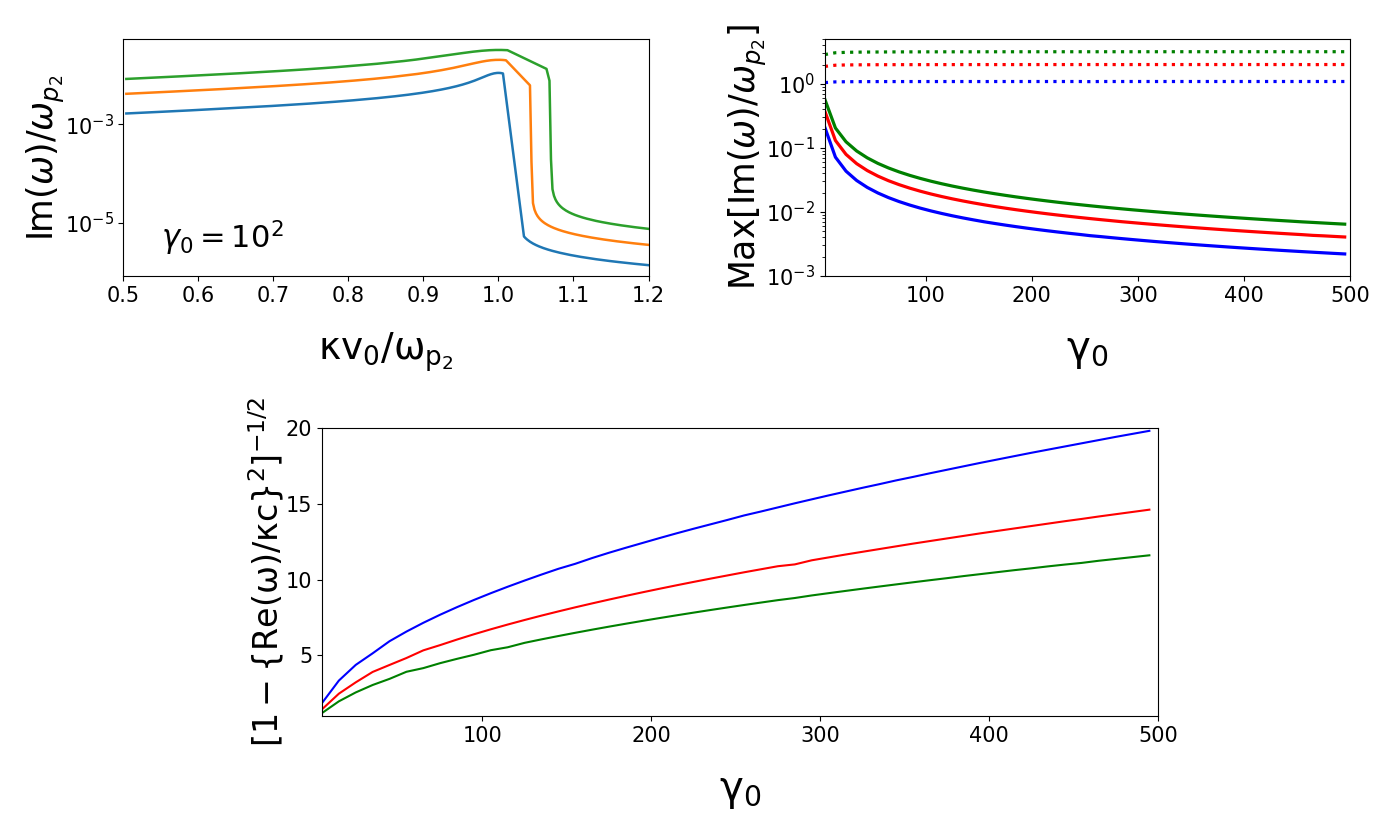}
\vskip -0.07in
\caption{Shown in the top left panel is the growth rate of the instability described in \S\ref{instability} as a function of wavenumber $\kappa$ for three different values of $\omega_{p_1}/\omega_{p_2}$, 2 (blue), 5 (red), 10 (green); $\omega_{p_2}$, given by eq. \ref{wp2}, is the plasma frequency of the $e^\pm$-medium encountered by the Alfven wave packet, $\omega_{p_1}$ is given by eq. \ref{wp1}, and $\kappa$ is the wavenumber for the instability. The growth rate of the instability is obtained by solving the dispersion eq. \ref{disersion-rel} to find the unstable mode. The top right panel shows the growth rate, -Im($\omega(k)$), of the fastest growing mode as a function of $\gamma_0$; $\gamma_0$ is the LF of e$^\pm$-plasma advected by the AW to supply the current density the wave needs, i.e. $c(\nabla\times {\bf B}_y)_z/4\pi$. The three different colored curves correspond to different $\omega_{p_1}/\omega_{p_2}$ as before. The maximum growth rate scales as $\gamma_0^{-1} \omega_{p_1}^{2/3}\omega_{p_2}^{1/3}$, as shown by the 3-dotted color curves which are -Im($\omega(\kappa)$)*$\gamma_0$. The bottom panel displays the effective Lorentz factor of the most unstable mode, i.e. $[1-({\rm Re}(\omega)/\kappa c)^2]^{-1/2}$, for the three different values of $\omega_{p_1}/\omega_{p_2}$, viz. 2, 5 \& 10.  }
\label{fig-growth-rate}
\end{figure*}

We have performed various checks to ensure that the results of numerical calculations presented in this work are reliable and accurate. Our simulation code conserves the total particle number of each charge sign, and the plasma \& displacement current-densities it calculates are consistent with the curl of magnetic field. We also checked for numerical stability and errors by changing the size of the spatial grid scale by a factor of 2--20 and found little change to the final results for the electric field, and particle speeds over a fairly long time baseline of $\sim 10^2 T_a$. Moreover, we cross-checked our numerical solutions with analytic results (eq. \ref{Ez-2}) for $t$ less than a few $T_a$ and found good agreement. And finally, the rate of exponential increase of $E_z$ on a longer time scale shown in Fig. \ref{fig-Ez2} (lower panel) agrees with the linear instability growth rate calculated in \S\ref{instability} and presented in Fig. \ref{fig-growth-rate}.

\subsubsection{A 2-stream like instability associated with CC-AW moving through stationary plasma}
\label{instability}

The starting point of linear stability analysis is to perturb equation (\ref{Ez-eq1}), which gives
\begin{equation}
    \partial^2_t E_z + \omega_{p_2}^2 E_z  = \partial_t \delta \left[ c(\nabla\times B)_z - 4\pi j_{z_1})\right] = -4\pi \partial_t (\delta j_{z_1}),
  \label{Ez-eq2}
\end{equation}
when the stationary plasma encountered by the CC-AW are accelerated to speed much less than $c$. We note that $E_z$ is already a first order perturbation variable, which is non-zero only when $\delta j_{z_1}\not=0$. This is why the perturbation to $\omega_{p_2}$ does not show up on the left side of the above equation. Moreover, the magnetic field $B_y$ associated with the AW is pre-specified (eq. \ref{By-tz}), and its perturbation is taken to be zero in the present analysis.

The equation for current density $j_{z_1}$ 
\begin{equation}
    \partial_t j_{z_1} = {q^2 n_1 E_z\over m \gamma_0^3} - \partial_z(v_1 j_{z_1}),
\end{equation}
-- which has the same form as eq. \ref{jz2-t-exact} -- is perturbed to obtain an equation for $\delta j_{z_1}$ 
\begin{equation}
    \partial_t \delta j_{z_1} + v_0 \partial_z(\delta j_{z_1}) = {q^2 n_1 E_z\over m \gamma_0^3} - j_{z_1} \partial_z \delta v_1 - \delta v_1 \partial_z j_{z_1},
\end{equation}
where $v_1=v_0 +\delta v_1$, and $v_0$ \& $\gamma_0$ are independent of $z$ as per the initial condition for CC-AW. Both $n_1$ and $j_{z_1}$ are functions of $z$ as given by equation (\ref{init-cond}), and they oscillate with time with frequency $\omega_a\ll\omega_{p_2}$ as follows
\begin{equation}
    j_{z_1}(z,t) = \epsilon B_0 k_x \cos(\omega_a z/c - \omega_a t),   \quad n_1 = {j_{z_1}\over q v_0 {\rm sign}(j_{z_1})}.
\end{equation}

The equation of motion of a charge particle
\begin{equation}
   {d v_1\over dt} = {q E_z\over m\gamma_0^3}
\end{equation}
is perturbed to give
\begin{equation}
   \partial_t \delta v_1 + v_0 \partial_z \delta v_1 = {q E_z\over m\gamma_0^3}
\end{equation}
Taking the $t$ \& $z$ dependence of perturbations $\delta v_1$, $\delta j_{z_1}$ and $E_z$ to be $\exp(i\omega t - i \kappa z)$, where 
$\kappa$ is the wavenumber of the instability, these equations reduce to
\begin{equation}
   i(\omega - v_0 \kappa) \delta j_{z_1} = {q^2 n_1 E_z\over m \gamma_0^3} + (i \kappa j_{z_1} - \partial_z j_{z_1})\delta v_1
\end{equation}
and
\begin{equation}
    i(\omega - v_0 \kappa)\delta v_1 =  {q E_z\over m \gamma_0^3}.
\end{equation}
Since $n_1$ and $j_{z_1}$ are functions of time, the above two equations are approximate and only valid when ${\rm Re}(\omega)\gg \omega_a$; we show below that ${\rm Re}(\omega)\approx \omega_{p_2} \gg \omega_a$. Eliminating $\delta v_1$ from these two equations we find
\begin{equation}
   i(\omega - v_0 \kappa)^2 \delta j_{z_1} =  {q^2 n_1 E_z\over m \gamma_0^3} \left[ \omega + i v_0\partial_z (\ln j_{z_1})\right],
\end{equation}
where we used $j_{z_1} = q n v_0$. 

Substituting this back into equation (\ref{Ez-eq2}) we obtain the dispersion relation
\begin{equation}
   \omega_{p_2}^2 - \omega^2 = -{4\pi q^2 n_1\omega\over m \gamma_0^3} \left[
 {\omega + i v_0\partial_z \ln j_{z_1} \over (\omega - v_0 \kappa)^2 }\right].
 \label{disersion-rel}
\end{equation}
One of the four roots of this equation has an imaginary component with negative sign corresponding to exponentially growing solution. The growth rate for a few different parameters are shown in Fig. \ref{fig-growth-rate}. Numerical solutions of equation (\ref{disersion-rel}) show that the growth-rate is highest for 
\begin{equation}
   \quad\quad {\rm Re}(\omega)\approx c \kappa\approx \omega_{p_2}, 
 \end{equation} 
and the dependence of this maximum growth rate on $\gamma_0$, $\omega_{p_1}$, and $\omega_{p_2}$ is (see the top right panel of Fig. \ref{fig-growth-rate})
\begin{equation}
    {\rm Max\bigl[|Im}(\omega)|\bigr] \approx {\omega_{p_1}^{2/3}\omega_{p_2}^{1/3} \over \gamma_0},
    \label{max-growth-rate}
\end{equation}

Since the instability described in this sub-section is really an overstable oscillation, i.e. Re($\omega)\not=0$, with finite wavenumber $\kappa$. A quantity of interest for later use in this work is the Lorentz factor with which the oscillations move forward, and that is defined as follows
\begin{equation}
    \gamma_{\rm os} \equiv {1\over \left[1-\left\{{\rm Re}(\omega)/\kappa c\right\}^2\right]^{1/2}}.
    \label{gam-os}
\end{equation}
The values of $\gamma_{\rm os}$ as a function of $\gamma_0$ for $\omega_{p_1}/\omega_{p_2} = 2$, 5 \& 10 are shown in the lower panel of Fig. \ref{fig-growth-rate}. For $\gamma_0\sim 10$, $\gamma_{\rm os}\sim 2$, and $\gamma_{\rm os}\propto \gamma_0^{1/2}$ is a fairly accurate description for how $\gamma_{\rm os}$ increases for $\gamma_0\gtrsim 10$ (see Fig. \ref{fig-growth-rate}).

We close this sub-section by providing a brief discussion of the physical origin of the instability. A key component, or trigger, for the instability is the current density, $j_{z_1}$, associated with charge particles that are being advected by the Alfv\'{e}n wave. A seed electric field $E_z$ perturbs this current density, $\delta j_{z_1} $, which in turn has a positive feedback on the electric field. The first term in equation (\ref{Ez-2}) is the important one for the feedback as it causes $\delta j_{z_1} $ to oscillate with frequency $\omega_{p_2}$, which then couples to the electric field resonantly. This feedback loop leads to an exponential increase with time of the component of the electric field $E_z$ that oscillates at frequency $\omega_{p_2}$, and thus comes to dominate the second term in equation (\ref{Ez-2}) after a few growth times as we see in Fig. \ref{fig-Ez2}.

\begin{figure*}
\centering
\hspace*{-0.5cm}
\includegraphics[width = 1.03\textwidth, height=11cm]{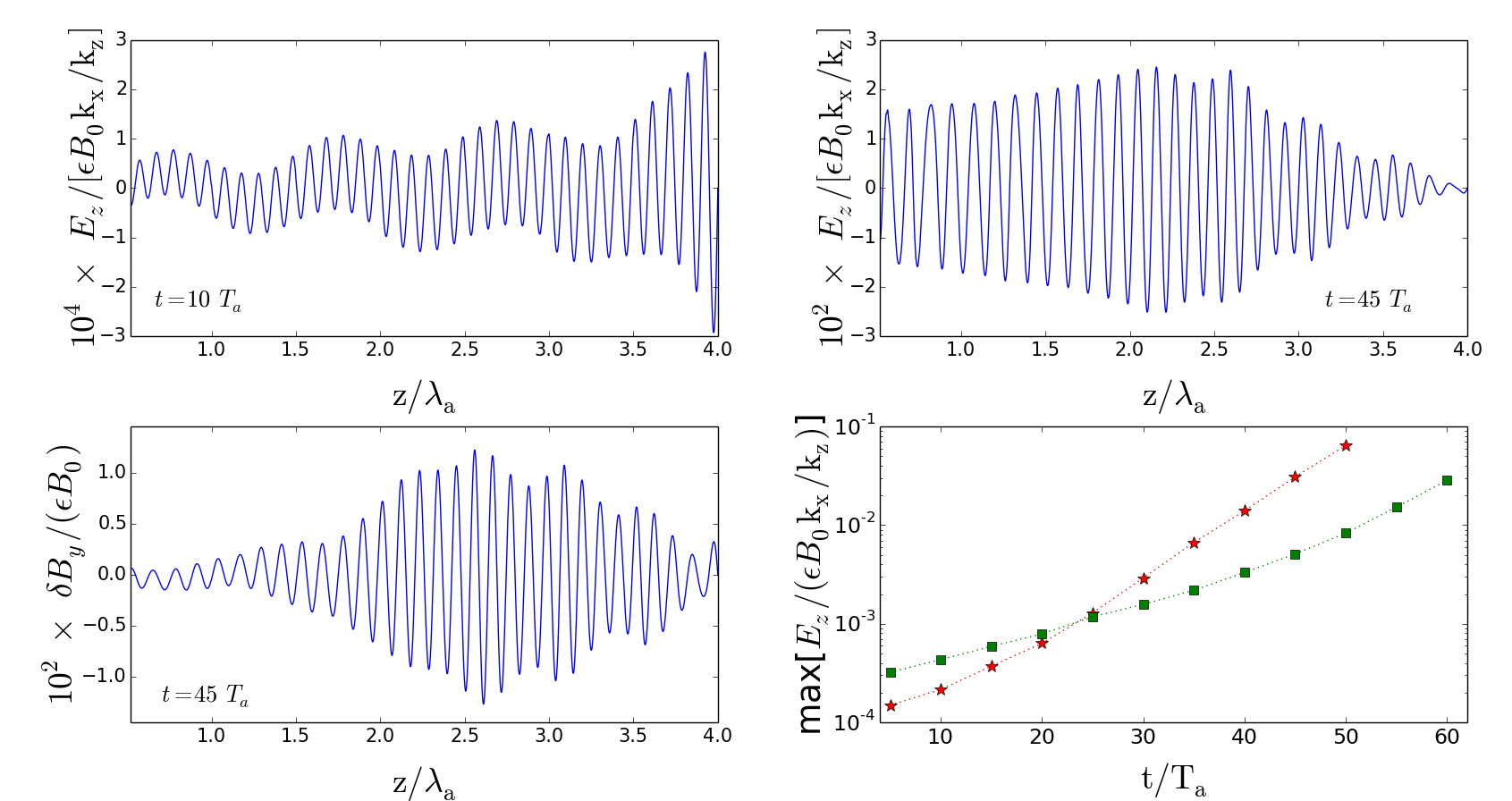}
\vskip -0.0in
\caption{Shown in the top left and right panels are the electric field strength divided by the Alfv\'{e}n wave amplitude, $E_z/(\epsilon B_0)$, at times $10 T_{a}$ and $45 T_{a}$ respectively and for $k_x/k_z=0.2$ -- the other parameters are the same as in the calculations presented in Figs. \ref{fig-vp-Ez} \& \ref{fig-Ez2}, i.e. $\gamma_0=10^2$, $\omega_B/\omega_a = qB_0(m c \omega_a) = 10^5$, $\omega_{p_2}/\omega_a =7.1$; $T_{a}=2\pi/\omega_a$ \& $\lambda_a = cT_a$ are Alfv\'{e}n wave period and wavelength respectively. Simulation results presented in this figure were calculated using the full set of equations \ref{EM-eqs}--\ref{particle-dynamics} without any simplifying assumption. These results should be contrasted with results in Figs. \ref{fig-vp-Ez} \& \ref{fig-Ez2} where the magnetic field perturbation ($B_y$) and the x-component of electric field perturbation associated with the AW packet ($E_x$) were prescribed for all $z$ \& $t$ and not calculated self-consistently as is done here.  The perturbation to Alfv\'{e}n wave amplitude, $\delta B_y$, -- which is the deviation of the y-component of the magnetic field from the expression given in eq. \ref{By-tz} -- is shown in the lower left panel. The magnitude of $\delta B_y$ is of the same order as $E_z$ (top right panel), and is identical to $\delta E_x$. The lower right panel shows the peak value of $E_z/(\epsilon B_0)$ as a function of time for two different values of $e^\pm$ plasma density encountered by the AW -- the density $n_2/{\rm max(n_{1})}$ for the points marked by red-stars is 0.5, whereas the density ratio is 0.125 for points denoted by green-squares; where $n_{1}$ is given by eq. \ref{init-cond}, and  all points in this panel were calculated using numerical simulations of Alfv\'{e}n wave propagation without any approximation. The growth rates of the instability we find here have roughly the same magnitudes as in Fig. \ref{fig-Ez2}.   }
\label{fig-Ez2-exact}
\end{figure*}

\subsection{Charge starved Alfv\'{e}n wave propagation: solution of exact equations}
\label{AW-subsec2}

In the previous sub-section (\S\ref{AW-subsec1}), we assumed that the magnetic field perturbation amplitude associated with the Alfv\'{e}n wave packet ($B_y$) does not evolve with time as the wave propagates out and an electric field along $z$ develops. This assumption is not strictly correct; it was introduced to simplify analytical calculations and gain physical understanding of the CC-AW system and the emergence of $E_z$. We drop that assumption in this sub-section and solve, numerically, the exact set of equations (\ref{EM-eqs}-\ref{EM-eqs2}) for the EM field. We keep track of three EM field components, $B_y$, $E_x$ and $E_z$, as the CC-AW propagates out and encounters a non-zero density, stationary, plasma. Equations for plasma density evolution and particle dynamics (\ref{particle-dyna0}) are unchanged. We continue to approximate the system as 1D in space by assuming that the spatial variation in the x direction is given by $\exp(i k_x x)$. This is a good assumption as long as $k_x \ll k_z$ and $k_x$ is not a function of $z$. The latter is guaranteed to be valid due to the symmetry of the system or the fact that {\bf B}$_0$ is independent of $x$ and thus $k_x$ is a conserved quantity. This can be seen from the following Eikonal equation
\begin{equation}
    {d k_i\over dt} = -c\sum_{j=1}^3 k_j {\partial \hat{B}_{0j}\over \partial x_i} \quad\implies\quad {d k_x\over dt} = 0.
\end{equation}

Results of numerical simulations of the full set of equations are presented in Fig. \ref{fig-Ez2-exact} at $t=10 T_a$ \& $45 T_a$ for the same set of parameters as in Figs. \ref{fig-vp-Ez} \& \ref{fig-Ez2} except that $ck_x/\omega_{p_2}=0.2$ for calculations in Fig. \ref{fig-Ez2-exact} as opposed to 0.5 in the other two figures. The structure of $E_z$ as a function of $z$ at different times with and without the approximation are similar -- both of which show pronounced oscillations at frequency $\omega_{p_2}$, which have similar amplitudes to within a factor $\sim 2$ after we correct for the difference in their respective $k_x$ values; $E_z\propto k_x$ (eq. \ref{EM-eqs}). Moreover, both calculations show a superposition of oscillations of frequencies $\omega_a$ and $\omega_{p_2}$ at early times ($t\lta20T_a$), but later on the exponentially growing mode at frequency $\omega_{p_2}$ becomes the dominant component.

The growth-rate of $E_z$ with time is shown in the bottom right panel of Fig. \ref{fig-Ez2-exact}. The exact calculations find growth-rates that are similar in magnitude to those of the approximate calculations shown in Fig. \ref{fig-Ez2} after we correct for the factor 2.5 difference in their $k_x$ values. Thus, the instability described in \S\ref{instability} survives intact for the full set of equations where the magnetic field perturbation, $B_y$, and the $\hat x$-component of the electric field ($E_x$) are accorded full dynamical status and are calculated self-consistently. The results we have presented in these figures are for $\gamma_0=10^2$. We have verified numerically that the results for the exact and approximate models are similar for $\gamma_0$ between about 10 and 200. One difference we found, however, is that the growth rate, Im($\omega$), declines as $\gamma_0^{-1}$ when $B_y$ and $E_x$ are pre-set and not allowed to be perturbed, and significantly faster when the dynamics of these fields are included in the calculation self consistently. Our numerical code for the exact solution of eqs. \ref{EM-eqs}--\ref{particle-dyna0} is unable to handle $\gamma_0\gtrsim 200$, and therefore we are unable to quantify the behavior of the instability at larger values of $\gamma_0$. However, according to the approximate analytical calculation of the growth rate presented in \S\ref{instability} when $B_y$ and $E_x$ are pre-specified, Im($\omega$) is exactly proportional to $\gamma_0^{-1}$ for all values of $\gamma_0$ larger than about 10. 

The perturbation to $B_y$ is shown in the lower left panel of Fig. \ref{fig-Ez2-exact}. This perturbation, $\delta B_y$, is defined to be the difference between the exact numerical solution $B_y(t,z)$ of eqs. \ref{EM-eqs}--\ref{particle-dyna0} and the approximate analytic solution given by equation (\ref{By-tz}). The magnitude of $\delta B_y$ is similar to $E_z$ (compare the top right and bottom left panels of fig. \ref{fig-Ez2-exact}) modulo the factor of $k_x$ for $E_z$. This means that $k_x \delta B_y$ is much smaller than the $\partial_t E_z\sim \omega_{p_2}E_z$ term in the second part of eq. (\ref{EM-eqs}) when $ck_x\ll\omega_{p_2}$, and that is the reason that the results we obtained when we neglected $\delta B_y$ did not make a large qualitative difference to the final result for $E_z$. The value of $ck_x/\omega_{p_2}$ for our numerical calculations is $\sim$0.05 whereas for Alfv\'{e}n waves in NS magnetosphere the expected value is $\sim 10^{-5}$. Our numerical code cannot handle $ck_x/\omega_{p_2}\lta 10^{-2}$, but based on the physics of the system we expect the growth-rate of $E_z$ presented in this work to apply to parameters appropriate for NS magnetospheres. We note that the electric field perturbation $\delta E_x$ is almost identical to $\delta B_y$, which is what one expects from the first part of eq. (\ref{EM-eqs}).

\section{Radiation from the coherent current oscillating at plasma frequency}

\begin{figure*}
\centering
\hspace*{-0.5cm}
\includegraphics[width = 0.93\textwidth]{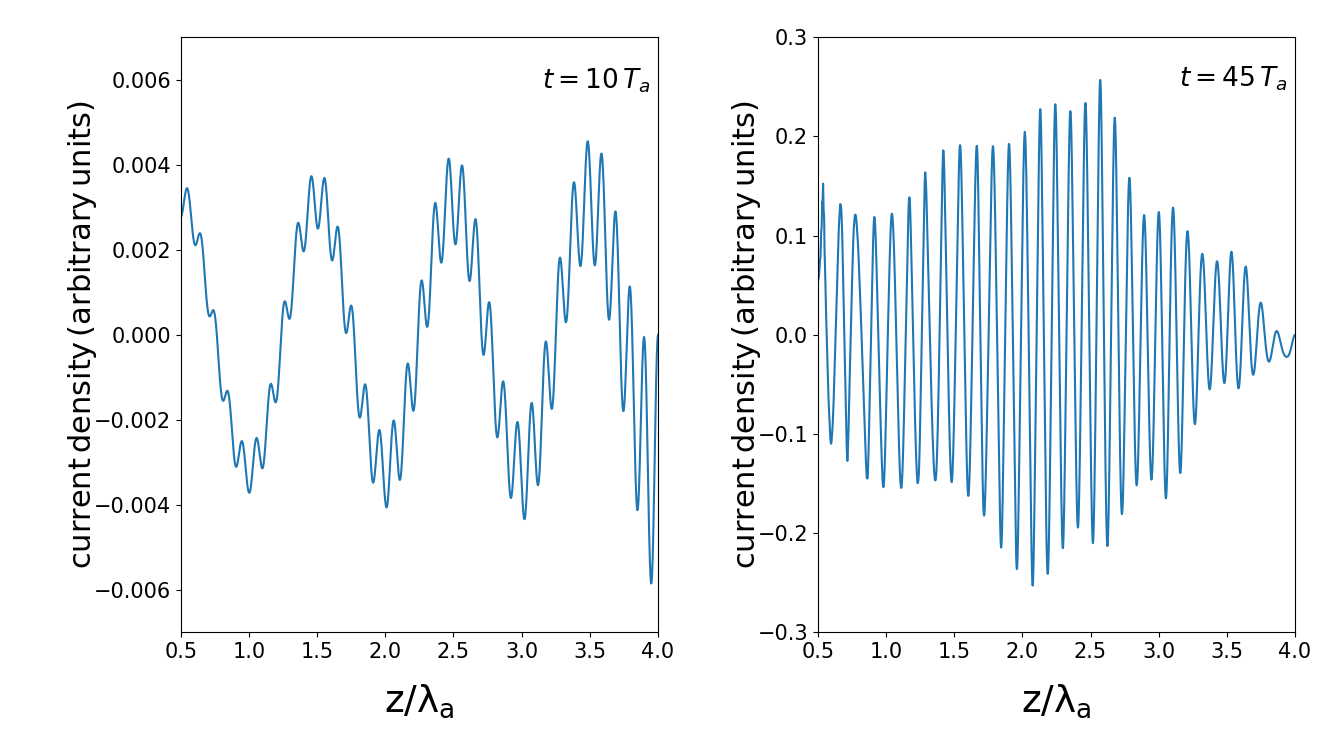}
\vskip -0.1in
\caption{Shown are current densities, $j\equiv q(n^+ v^+ - n^- v^-)$, at times $10 T_{a}$ (left panel) and $45 T_{a}$ (right panel).
The parameters for these calculations are the same as in Fig. \ref{fig-Ez2-exact}.
Simulation results presented in this figure were calculated using the full set of equations \ref{EM-eqs}--\ref{particle-dynamics} without any simplifying assumption. The current density at $t=10 T_a$, when the perturbation in linear, is sum of two sinusoidal functions of frequencies $\omega_a$ and $\omega_p$ (Alfven wave and plasma frequencies respectively). At $45 T_a$, the perturbations are mildly nonlinear with particle velocity $\sim 0.3$c, and the current density has fluctuations on roughly the plasma length scale $c/\omega_p^{-1}$.
}
\label{fig-jp}
\end{figure*}

The oscillating current associated with the instability described in \S\ref{instability}, see Fig. \ref{fig-jp}, produces coherent radiation. The starting point of the calculation of the emergent luminosity is the following exact wave equation
\begin{equation}
    \nabla^2{\bf B} - \frac{1}{c^2}\nabla_t^2 {\bf B} = - \frac{4\pi}{c} \nabla\times{\bf j} \equiv -4\pi {\bf S},
\end{equation}
where ${\bf j}$ is the current density induced by the instability, and {\bf S} is the source for EM waves generated by the oscillating current density. The solution of this equation is
\begin{equation}
\begin{split}
    {\bf B(r},t) & = -\int d^3r_1\, { {\bf S(r_1}, t - |{\bf r - r_1}|/c)\over |{\bf r - r_1}|} \\
    & = -{1\over c} \nabla\times \int d^3r_1\, { {\bf j(r_1}, t - r/c + {\bf \hat n\cdot r_1}/c)\over r}
\end{split}
\label{Bw1}
\end{equation}
where $\mathbf{r}$ is the location at which the field is measured, $\mathbf{r_1}$ is the location of the oscillating charge at the retarded time $t_1 = t-r/c+(\mathbf{\hat n}\cdot\mathbf{r_1})/c$, and 
$\mathbf{\hat n}\equiv(\mathbf{r}-\mathbf{r_1})/\vert \mathbf{r}-\mathbf{r_1}\vert$ is the unit vector that points 
from the latter to the former location. As usual, the expansion $|{\bf r - r_1}| \approx r - {\bf\hat n\cdot r}_1$ far away from the source ($r\gg r_1$) has been used.

The current density oscillates along the wave propagation direction ($z$) at a plasma frequency $\omega_p$ (Fig. \ref{fig-jp}), but with amplitude that varies along $z$. It varies along the other two directions on a much longer length scale of $\sim k_x^{-1}$. Let us write the functional form of ${\bf j}$ as
\begin{equation}
    {\bf j}({\bf r}_1, t_1) = j_0 {\bf \hat z}\; \xi(x_1, y_1)\; \exp(ik_p \eta - \eta^2/\ell_0^2),\quad \eta\equiv z_1 - v t_1,
\end{equation}
where $v$ is the pattern speed of the oscillating current.
Substituting this in equation (\ref{Bw1}) we find
\begin{equation}
 \begin{split}
    {\bf B(r},t) & = {i j_0 k \, {\bf \hat z\times \hat r} \, \exp(-i\omega t + i k r)\over c r} \int dx_1 dy_1 \xi(x_1, y_1) \; \\ %removed the minus sign (RG)
   & \times\;\int dz_1 \exp\left[ i k_p z_1 - i k {\bf\hat n\cdot r_1} - (z_1 - \beta {\bf\hat n\cdot r_1})^2/\ell_0^2\right]
    \end{split}
\end{equation}
where 
%$k \equiv k_p v/c$. 
$k=\beta k_p$ and $\beta = v/c$. 
Taking ${\bf\hat n = \hat x} \sin\theta_0 + {\bf\hat z} \cos\theta_0$, without loss of generality, and using the 
small $\theta_0$ and large $\gamma$ approximations, 
so that $(1-\beta\cos\theta_0)\approx(1+\gamma^2\theta_0^2)/2\gamma^2$, results in
\begin{equation}
\begin{split}
   & {\bf B_\perp(r},t) = {\bf A}_B \int dx_1 dy_1 \xi(x_1, y_1)\; \exp(-i k x_1 \theta_0) \; \times \\
    &\quad \int dz_1 \, \exp\left[ -\left\{ {z_1(1 +\theta_0^2\gamma^2)\over 2\gamma^2 \ell_0} - {\beta x_1\theta_0\over \ell_0}\right\}^2 + {i k_p z_1 (1 +\theta_0^2\gamma^2)\over 2\gamma^2}\right],
    \end{split}
    \label{B-perp3}
\end{equation}
where ${\bf A}_B = ({\bf \hat z\times \hat r}) \,k j_0\, \exp(-i\omega t + i k r) /cr$
and 
%$\beta=v/c$, 
$\gamma = (1 - \beta^2)^{-1/2}$. It is convenient to rescale, $z_1' = z_1 (1 +\theta_0^2\gamma^2)/2\gamma^2$, and shift the $z'_1$ 
integral range by $\beta x_1\theta_0$ to simplify the above expression:
\begin{equation}
 {B_\perp(r},t) = {2 \gamma^2 A_B\over 1 + \gamma^2\theta_0^2} \int dx_1 dy_1\, \xi(x_1, y_1) % the bold face of B was removed since we only want the magnitude (RG)
   \int dz'_1 \, \exp\left[ -{z_1'^2\over \ell_0^2} + i k_p z'_1\right],
\end{equation}
If the domain of $z_1'$ integration were to be [$-\infty, \infty]$, then
\begin{equation}
   {B_\perp(r},t) = {2\pi^{1/2} \gamma^2 A_B \ell_0 \exp(-k_p^2 \ell_0^2/4)\over 1 + \gamma^2\theta_0^2} \int dx_1 dy_1 \xi(x_1, y_1)
\end{equation}

Let us take $\ell_{\perp\rm coh}$ to be the smaller of the transverse wavelength of the AW and the size of the transverse coherent emission region. The amplitude of the wave magnetic field then is
\begin{equation}
   {B_\perp(r},t) \approx {2\pi^{1/2} \gamma^2  k j_0 \ell_0 \ell_{\perp \rm coh}^2 \exp(-k_p^2 \ell_0^2/4)\over (1 + \gamma^2\theta_0^2)\, r c},
   \label{B-perp4}
\end{equation}
and the luminosity when $\lambda/\ell_{\perp \rm coh} < \gamma^{-1}$ and $\gamma\theta_0\approx1$ is
\begin{equation}
   L = r^2 B_\perp^2 c \approx {\pi \gamma^4  k^2 j_0^2 \ell_0^2 \ell_{\perp \rm coh}^4 \exp(-k^2 \ell_0^2/2)\over c},
   \label{Lum1}
\end{equation}
where for $\gamma\gg1$, $\beta\sim1$ and $k_p\approx k$. The exponential suppression factors in equations \ref{B-perp3} \& \ref{Lum1} are there only when the length of the oscillating current is $\gg\ell_0 \gamma^2$ and its temporal and spacial structure is sinusoidal. Otherwise, the luminosity is much larger than given by eq. \ref{Lum1}. 
The luminosity in (\ref{Lum1}) is the isotropic equivalent in the observer frame except for the cosmological redshift factors, and it is beamed within an angle $\min\{\gamma^{-1}, \lambda/\ell_{\perp \rm coh}\}$; $\lambda=2\pi/k$ is the wavelength of the observed radiation. 

For a Gaussian envelope of the sinusoidally oscillating current density with $\ell_0 \gg k^{-1}$, the exponential suppression factor in equation (\ref{Lum1}) makes the radiative loss too small to be of any  practical importance. The current density during the linear growth phase of the instability has the sinusoidal shape with amplitude that varies over a few $\lambda_a$ as shown in Fig. \ref{fig-jp}, and therefore radiative losses are small during this period.

As the instability grows and enters the non-linear regime, i.e. when the $e^\pm$ speed becomes close to $c$, the current density fluctuations increase rapidly in space and time, and radiative losses are not small any more. The current density in a segment of plasma of length $\sim k_p^{-1}$ decoheres after the plasma has traveled a distance of order $\ell_{dc} \sim \gamma^2 k_p^{-1}$. This is expected from causality as the length in the frame moving with the pattern speed is larger by a factor $\gamma$ and the elapsed time in this frame is smaller than the NS rest frame also by a factor $\gamma$; this result is also supported by our simulations. In this regime, radiation from different regions of size $\ell_{dc}$ can be added incoherently\footnote{For a purely sinusoidal envelope, of extent much larger than $\gamma^2/k$, the bandwidth of the emergent frequency is very narrow.}, and the resulting luminosity from the length of the Alfven wave packet where the instability has entered the nonlinear phase ($z_{cs}$), is $z_{cs}/\ell_{dc}$ times\footnote{To see that the luminosity is larger by a factor $z_{cs}/\ell_{dc}$ instead of $z_{cs} k_p$ it is best to go back  to eq. (\ref{B-perp3}). When the phase term in the exponent of that integrand is coherent over length scale $\gamma^2 k_p^{-1}$ -- which is the distance traveled by the wave when the signal from one end of a region of size $k_p^{-1}$ travels to the other end -- the contribution to the integral from that is $A_B \gamma^2 k_p^{-1} \ell_{\perp coh}^2$. Adding different segments of length $\gamma^2 k_p^{-1}$ incoherently gives the radiated power to scale as $(cr^2) (A_B \gamma^2 k_p^{-1})^2 [z_{cs}/(\gamma^2 k_p^{-1})$]. } the luminosity given by equation (\ref{Lum1}) with $k\ell_0 \sim 1$.
\begin{equation}
    L \sim {\gamma^4  j_0^2 \ell_{\perp \rm coh}^4 z_{cs}\over \ell_{dc} c}\sim {\gamma^2  j_0^2 \ell_{\perp \rm coh}^4 k\, z_{cs}\over c}
   \label{Lum2}
\end{equation}
If the transverse size of the region which contributes to the observed radiation is $\ell_{\perp} > \ell_{\perp \rm coh}$, then the above expression for luminosity should be multiplied with a factor $(\ell_\perp/\ell_{\perp \rm coh})^2$.

The particle density for plasma frequency $\sim 10^9$Hz is $n_e\sim 10^{10}$ cm$^{-3}$. The current density associated with the instability, 
during the non-linear phase, is $j_0\sim q n_e c\sim 2{\rm x}10^{11}$ cgs. For $\ell_{\perp\rm coh} \sim 10^4$cm, $\gamma\sim 10$, and $z_{cs}\sim 10^8$cm, 
the emergent isotropic equivalent luminosity, due to the fluctuating current associated with the instability is $\sim 10^{38}$ erg s$^{-1}$, and for $\omega_p/2\pi\sim 10$ GHz, $L\sim 10^{42}$ erg s$^{-1}$. 
The radiation is beamed in a narrow cone of angle $\lambda/\ell_\perp\sim 3{\rm x}10^{-3}$rad, and its frequency is of order $0.1 \omega_p$ as the 2-stream instability growth rate peaks at at a wave number $\sim 0.1 k_p$. 
%For $\ell_\perp/\ell_{\perp \rm coh}\sim 10$, the emergent luminosity is on the order of a typical cosmological FRB. 

%For a larger $\gamma$ by a factor of a few, the radio luminosity is of order what we observe for a typical cosmological FRB.

The rate of loss of energy per particle due to this radiation is $p_r \sim L(\lambda/\ell_{\perp \rm coh})^2 \gamma^{-2}/[n_e \ell_{\perp \rm coh}^2 z_{cs} ]\sim m_e c^2 \omega_p$; the factor $\gamma^{-2}$ in the expression for $p_r$ takes into account that the radiation emitted in NS rest frame in time interval $\delta t$ arrives at the observer over a smaller time duration of $\delta t/\gamma^2$. Thus, the radiative loss time is $\sim \omega_p^{-1}$ when particles are not accelerated to highly relativistic speed by the instability. 

The shortest growth time for the instability is of order $ \omega_p^{-1}$ when the relative speed of counter-streaming particles is mildly relativistic (Fig. \ref{fig-growth-rate}). As discussed before, the radiative loss time is much larger than the plasma time during the linear growth phase of the instability (due to the exponential suppression factor in eq. \ref{Lum1}), and therefore the instability grows rapidly on the plasma timescale. When the speed of $e^\pm$ approaches $c$ and the plasma becomes turbulent, and radiative losses increase dramatically as given by eq. (\ref{Lum2}), and the instability saturates at that point. 

This result can be rephrased in the following way. The EM radiation comes from plasma modes moving at the pattern speed with Lorentz factor $\gamma$ (which is of order $\gamma_0^{1/2}$). As long as the rate at which energy is being pumped into these modes is faster than that being radiated away, the modes continue to grow. The total energy density contained in the modes is $E_z^2/(4\pi)$, which can be shown to be $\sim ~ n m c^2$.  When $E_z$ reaches the strength such that the non-linearity parameter $E_z q/ (m \omega_p c) \sim 1$, radiative losses become of order the energy pumped into the modes by the instability and further growth of plasma mode amplitudes is terminated.

\section{Interaction of charge starved Alfv\'{e}n waves moving in opposite directions}
\label{AW-colliding}

Alfv\'{e}n waves moving in opposite directions along a magnetic field line is a likely scenario when the neutron star magnetosphere is shook up by a crustal disturbance. This is because magnetic field perturbations are launched at the surface of the NS where these field lines are anchored. These perturbations travel along the field lines away from the NS and they collide and interact somewhere in the magnetosphere. The interaction is particularly complex, and interesting, when it takes place in the charge starvation region for these Alfv\'{e}n waves and when the polarization angles of the counter streaming Alfv\'{e}n waves are different. 

The basic picture that is being suggested is that when Alfv\'{e}n waves are launched at opposite ends of a magnetic field bundle anchored on NS surface, 
these waves collide at some height above the NS surface. There are two possibilities to consider for the interaction between these Alfv\'{e}n waves moving 
in opposite directions. One of which is where the plasma density is marginally above the critical value for charge starvation everywhere along the wave 
trajectory when one considers the Alfv\'{e}n packet traveling from one end of the magnetic bundle to the other and ignore the wave packet coming from the 
other end of the magnetic bundle. However, the plasma density along the trajectory is insufficient to support the superposition of these two waves. 
Therefore, the system becomes charge starved in the region where these two waves collide, and it is unavoidable that a strong electric field parallel to 
the static magnetic field will develop rapidly and with strength of order the Alfv\'{e}n wave amplitude. Since the plasma density is sub-critical for the 
combined Alfv\'{e}n waves moving in opposite directions, they cannot advect charge particles with them to prevent charge starvation. This is because charge 
particles advected by the wave from a patch in the collision region would create a higher charge deficit there. The strong electric field in the collision 
region would accelerate charge particles, and under the right conditions they would generate coherent EM emission \citep{Kumar+17,KumarBosnjak2020, LuKumarZhang20, zhang20}.

The other possibility for the interaction between the counter-moving Alfv\'{e}n waves is that at least one of them has become charge starved at some height before running into the other wave. In this case, the ``charge starved'' wave is likely to advect charge particles with it as it travel further away from the NS. The CC-AW develops particle density fluctuation and electric field parallel to the magnetic field that also fluctuates on the plasma length scale (\S\ref{AW-numerics}). When the CC-AWs moving in opposite directions along the magnetic field bundle collide, strong two stream instability leads to formation of particle clumps and development of strong electric field parallel to the unperturbed magnetic field, and likely generation of strong coherent radiation. A discussion of the detailed physics of this interaction is outside the scope of this work.

\section{Discussion and Conclusion}

An Alfv\'{e}n wave packet that travels through a medium of ever decreasing plasma density will eventually become charge starved, i.e. it will find itself in a region where the charge density is too small to be able to supply the current needed by the wave even when particles are accelerated to the speed of light. The transition region where particles are rapidly accelerated and carried with the Alfv\'{e}n wave is not explored in this work. We assumed, instead, that after crossing this transition region, the wave advects just the sufficient number of particles with it, which move with Lorentz factor $\gamma_0\gg1$, to avoid charge starvation at larger radii. We have analyzed how the system of charge particles carried with the Alfv\'{e}n wave (CC-AW) interacts with plasma of finite density it encounters beyond the transition region, and the evolution of the system. The main result we find is that this interaction leads to an instability which is similar to the well known 2-stream instability. Particles advected by the AW, as well as those encountered by the CC-AW beyond the transition region, form clumps as a result of the instability, and a strong electric field along the direction of the unperturbed magnetic field ($E_z$) develops\footnote{\cite{Chen&Beloborodov2020} reported evidence for a weak 2-stream instability in their PIC simulations. This was likely due to the fact that they considered the Alfv\'{e}n wave after picking up charges to be moving into very low density medium -- almost vacuum, relatively speaking -- so that the growth rate of the instability was weak. Furthermore, they followed the Alfv\'{e}n wave propagation for only a few AW wavelength, and that is too short a time to see the development of a strong electric field and particle clump formation we find.}. The characteristic wavelength and growth time of the instability are the plasma length and frequency of the stationary plasma encountered by the CC-AW. The spatial scale for the instability is much smaller than the Alfv\'{e}n-wave wavelength ($\lambda_a$) by a factor $\gtrsim 10^4$ in neutron star magnetosphere, and thus according to our numerical simulations strong $E_z$ of order a few percent of AW amplitude develops after the CC-AW has traveled a distance of a few 10s of $\lambda_a$. 

The scenario we have analyzed in this work (CC-AW) is plausible under some physical situation depending on how the AW makes a transition to the sub-critical density medium; this scenario, where the AW picks up and transports charge particles with it at high LF, is suggested by the work of \cite{Chen&Beloborodov2020}. One of the main findings of this work is that development of strong $E_z$ even in this scenario where the AW is never truly charge-starved is unavoidable and therefore coherent radio emission should be generated. There is another possible scenario where plasma in the transition zone becomes clumped due to two-stream instability and a strong and oscillating electric field develops that accelerates these clumps to high speed. This too leads to coherent radio waves. We close the paper with a brief discussion of this possibility.

Electrons and positrons in the transition zone have speed close to that of light. The counter-streaming $e^\pm$ that supply the current to the AW are subject to the 2-stream instability. Even before the onset of the charge starvation, an electric field $E_z$, along the unperturbed magnetic field {\bf B}$_0$, develops on spatial scales of $c\omega_p^{-1}$ due to this instability; $\omega_p$ is the plasma frequency in the transition zone where the AW is starting to become 
charge starved. The strength of the field is of order $E_z\sim \delta j/\omega_p\sim \eta c B_y k_x/(4\pi\omega_p)$; where $B_y$ is the Alfv\'{e}n wave amplitude, 
and $\eta \equiv [1- |(4\pi j/c)|/(B_y k_x)]$ is the dimensionless plasma current deficit.  This electric field propagates outward like a traveling wave, as discussed 
in \S\ref{instability}, and therefore particles moving in the same direction as the electric-wave are accelerated for a time, $T_E$, that is longer than 
$\omega_p^{-1}$ by a factor that depends on the LF of the electric-wave front which is of order a few as shown in Fig. \ref{fig-growth-rate}. Thus, 
$T_E\sim 10 \omega_p^{-1}$, and particle LF $\gamma_0\sim 10 \omega_p^{-1} qE_z/(mc)\sim \eta\epsilon \omega_B \omega_a/\omega_p^2$; where 
$\epsilon\equiv B_y/B_0\sim 10^{-3}$ is the dimensionless Alfv\'{e}n wave amplitude, and $\omega_B \equiv q B_0/(mc)\sim 3{\rm x}10^{19}$rad s$^{-1}$ is the cyclotron frequency 
for magnetars at $\sim 10 R_{ns}$. Taking AW frequency $\nu_a\sim 10^4$Hz and $\nu_p\sim 10^{9}$Hz, we find $\gamma_0\sim 10^2$ when 
the charge deficit $\eta$ is of order unity. The clumps of particles moving along curved magnetic field lines would produce coherent emission which would also 
limit their LF to $\sim 10^2$. These clumps would last for at least the light-crossing time in the comoving frame of the clump and that is sufficient for the 
generation of coherent radiation.

\section*{Acknowledgments}

This work has been funded in part by an NSF grant AST-2009619. WL was supported by the Lyman Spitzer, Jr. Fellowship at Princeton University. Some of the work presented here was carried out while PK was visiting the Yukawa Institute (YITP), Kyoto. He is grateful to Prof. Kunihito Ioka for the hospitality, for organizing an FRB workshop during that visit, and for many stimulating science discussions. He would like to thank YITP for the  financial support provided under the Visitors' Program of FY2022, and YITP-W-22-18 for funds for the worshop. 

\bigskip\bigskip

\noindent {\bf DATA AVAILABILITY}

\medskip
The code developed to perform calculations in this paper is available upon request.

%\bibliographystyle{mnras}
%\bibliography{FRB} % if your bibtex file is called example.bib

\end{document}